\documentclass{ws-ijmpd}
\usepackage[super,compress]{cite}
\usepackage[driverfallback=dvips,breaklinks]{hyperref}
\hypersetup{colorlinks,urlcolor=black,citecolor=black,linkcolor=black,filecolor=black}
\usepackage{breakurl}

\begin{document}
	
	\thispagestyle{plain}

    \def\bib{B\kern-.05em{I}\kern-.025em{B}\kern-.08em}
    \def\btex{B\kern-.05em{I}\kern-.025em{B}\kern-.08em\TeX}

    \title{MULTIPLE RINGS IN THE SHADOW OF EXTREMELY COMPACT OBJECTS}

    \author{Jingkai Wang}
    \address{Sendelta International Academy, Tanghe 2nd Road\\Bao'an,Shenzhen, Guangdong, 518000, China\\vanscii@outlook.com}
    \maketitle
	\begin{abstract}
	The Event Horizon Telescope's image of the M87 black hole provides an exciting opportunity to study black hole physics.  Since a black hole's event horizon absorbs all electromagnetic waves, it is difficult to actively probe the horizon's existence. However, with the help of a family of extremely compact, horizon-less objects, named ``gravastars'', whose external space-times nearly identical to those of black holes, one can test the absence of event horizons: absences of additional features that arise due to the existence of the gravastar, or its surface, can be used as quantitative evidence for black holes. We apply Gralla {\it et al.}'s approach of studying black hole images to study the images of two types of gravastars: transparent ones and reflective ones. In both cases, the transmission of rays through gravastars, or their reflections on their surfaces, lead to more rings in their images. For simple emission models, where the redshifted emissivity of the disk is peaked at a particular radius $r_{\rm peak}$, the position of a series of {\it rings} can be related in a simple manner to light ray propagation: a ring shows up around impact parameter $b$ whenever rays incident from infinity at $b$ intersects the disk at $r_{\rm peak}$.   We show that additional rings will appear in the images of transparent and reflective gravastars.  In particular, one of the additional rings for the reflective gravastar is due to the prompt reflection of light on the gravastar surface, and appears to be well separated from the others.  This can be a robust feature, which may be reliably used to constrain the reflectivity of the black-hole's horizon. 
	\end{abstract}
	
	\section{Introduction}
	
	The Event Horizon Telescope produced the first ``image'' of the supermassive black hole at the center of the M87 galaxy, providing an exciting opportunity to study black hole physics. \cite{akiyama2019first,akiyama2019second,akiyama2019third,akiyama2019fourth,akiyama2019fifth, akiyama2019sixth}.  The ``shadow" and ``rings'' in the images, caused by the black hole's deflection of light emitted by the accretion disk, have been used to study various aspects of astrophysics and cosmology.  For example, the shadow of black holes with different redshift can be used to constrain Hubble constant and study cosmic expansion history \cite{Tsupko2019} as well as to test the cosmic expansion\cite{Perlick2018}. The images have also been applied to test general relativity~\cite{Ayzenberg2018,Psaltis_2015,bambi2009apparent,Psaltis2018,Johannsen2010}, and search for dark matter~\cite{Konoplya2019,Hou2018a,Hou2018}. 
	
	In Ref.~\cite{gralla2019black}, Gralla {\it et al.} analytically studied the images of a think disk surrounding a Schwarzschild black hole, using various models for disk emission and opacity.  They later extended their studies to spinning black holes \cite{Gralla2017}. In these works, the authors showed that {\it photon} and {\it lensing rings} are the generic features of these images. 
	
	The name ``Event Horizon Telescope'' indicates that we are probing the event horizon, the very defining feature of a black hole~\cite{misner1973gravitation}.  In this paper, we will focus on testing the existence of the horizon --- by studying features of shadows of objects that {\it do not} have horizons, and how they differ from the shadows of black holes. One class of such objects is the so-called {\it gravastars}: extremely compact objects whose external space-time geometry are well approximated by those of black holes, until very close to the horizon, where the vacuum space-time transitions into an object with a surface. Gravitational waveforms from gravastars differ from those from black holes, and such difference has been proposed as a way to search for gravastars, or to quantify how ``black'' the black holes really are~\cite{Nakao2018, Chirenti2007,pani2009gravitational,pani2010gravitational,cardoso2017tests}. Some authors have also proposed that gravastars give rise to different features in gravitational lensing ~\cite{2016Gravitational}.

	One family of gravastars was proposed by  Mazur and Mottola \cite{mazur2004gravitational}. In these gravastars, the external Schwarzschild black hole space-time transitions into an internal de-Sitter space-time at an infinitesimally thin spherical surface, which has a finite amount of surface tension and surface energy. 
	Propagation of null rays in the Mazur-Mottola gravastar space-time, and its shadows, assuming that it is transparent, have been studied by Ref.~\cite{sakai2014gravastar}. 
	In this paper, we will further apply the disk model from Ref. \cite{gralla2019black} to the gravastar.  We shall also consider an alternative model where the gravastar is reflective, which gives rise to different shadows. In situations where the disk as a simple emission profile --- namely one that has a peak observed intensity --- we also relate the appearance of multiple rings to the multiple ways light from the disk can travel toward the observer at infinity. 
	
	This paper is organized as follows. In Sec. \ref{sec:nullray}, we will discuss propagations of null geodesics in black hole and gravastar space-times. In Sec. \ref{sec:shadow}, we will discuss how the shadow profile is calculated and the disk model we will use. In Sec. \ref{sec:shadowdiffer}, we will compare differences between shadows of black holes and the two types of gravastars --- and relate these differences to the different ways light travel in these space-times. In Sec. \ref{sec:3c}, we will discuss some features of the rings and in Sec. \ref{sec:conclusion} we summarize our main conclusions.

	\section{Ray-Tracing and Disk Models}

	In this section, we review the equations of motion of null rays in the (Schwarzschild) black hole and gravastar space-times, discuss features of null geodesic motion, and introduce disk models considered by Ref.~\cite{gralla2019black}.

	\label{sec:nullray}
	\begin{figure*}[!t]
	\centering
		\includegraphics[width=\textwidth]{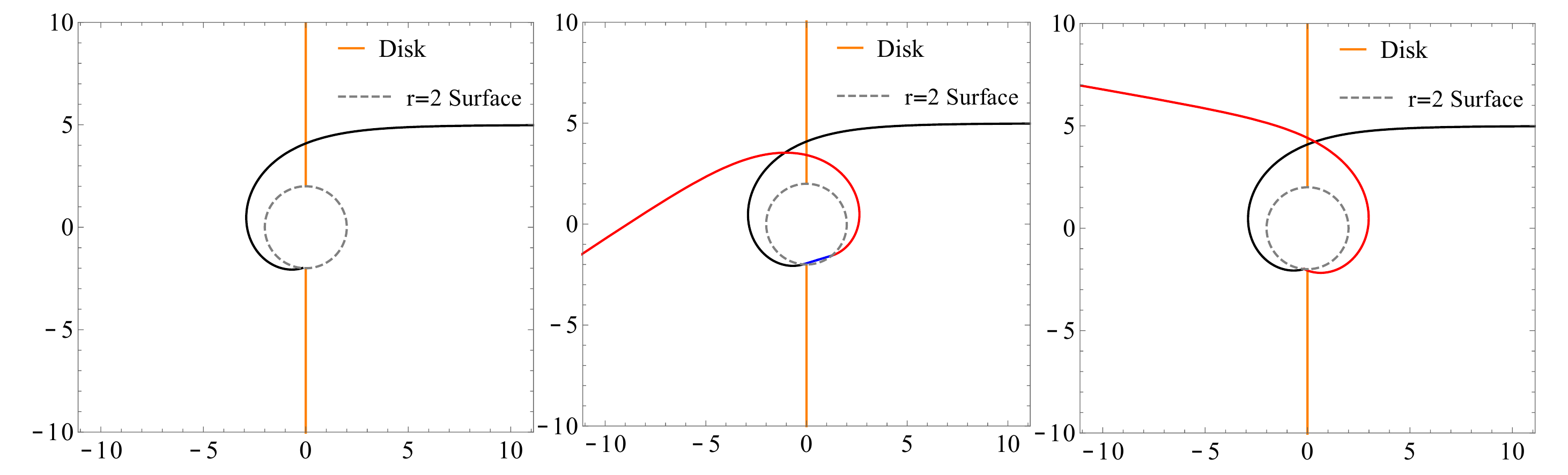}
		\caption{
		We plot the trajectory of null rays in the $(r\cos\phi,r\sin\phi)$ Cartesian coordinate system; the rays all enter from $\phi=0$ direction, with impact parameter $b=5M$; we also choose $a\approx 2M$ for the gravastars (see comment at the end of Sec.~\ref{sec:nullray}). The dashed gray line is $a=2$ surface where we assume the event horizon will locate. We use orange line segments (consisting $r>2M$ and $\phi=\pi/2$ or $\phi=3\pi/2$) to indicate the disk. The left panel shows the null ray trajectory for a black hole: the incoming ray (black color) intersects the disk once before reaching the horizon at $r=2M$. 
		The middle panel shows the null ray trajectory for a transparent gravastar; the ray first penetrates the gravastar, follows a straight trajectory (blue color), and then leaves the gravastar --- with the returning ray (red color) making one more intersections with the disk before returning to infinity.  The right panel shows the orbit for a reflective gravastar; after reflecting off the gravastar surface, the returning ray  (red color) intersects the disk two more time.}
		\label{compareorbit}
	\end{figure*}
	\subsection{Model of Gravastars and Null Geodesics}
	Metric of a Schwarzschild black hole and that of a gravastar can both be written in the same form: 
	\begin{equation}
	ds^2 =-f(r) dt^2 +\frac{1}{h(r)}dr^2+r^2(d\theta^2 +\sin^2\theta d\phi^2)\,.
	\end{equation}
	We have 	
	\begin{equation}
	    f(r) = h(r) = 1-\frac{2M}{r}\,, \quad r>2M\,,
	\end{equation}
    for  a black hole, and 
	\begin{equation}
	f(r) =
	\left\{
	\begin{array}{ll}
	\displaystyle h(r)=1-\frac{2M}{r}\,, &  r>a \\
	\\
	\displaystyle \alpha h(r) = \alpha\left( 1-\frac{8\pi\rho r^2}{3} \right) \,,\quad & 0< r<a
	\end{array}
	\right.
	\label{fcondi}
	\end{equation}
	for a gravastar. 	The value of $\alpha$ is fixed in such a way that $f(r)$ is continuous at $r=a$:
	\begin{equation}
	\alpha =\frac{\displaystyle 1-\frac{2M}{a}}{\displaystyle 1-\frac{8\pi\rho a^2}{3}}\,.
	\end{equation} 
	Here, one defines
	\begin{equation}
	M_v =\frac{4\pi \rho a^3}{3}
	\end{equation}
	which is the gravitational mass of the interior region of the gravastar $r<a$.  There is a shell located at $r=a$, which supplies energy and stress in order for Einstein's equation to be satisfied in the vicinity of the shell.  In the Mazur-Mottola case, $M_v = M$.

	\begin{figure*}[t]
	\centering
    \includegraphics[width=\textwidth]{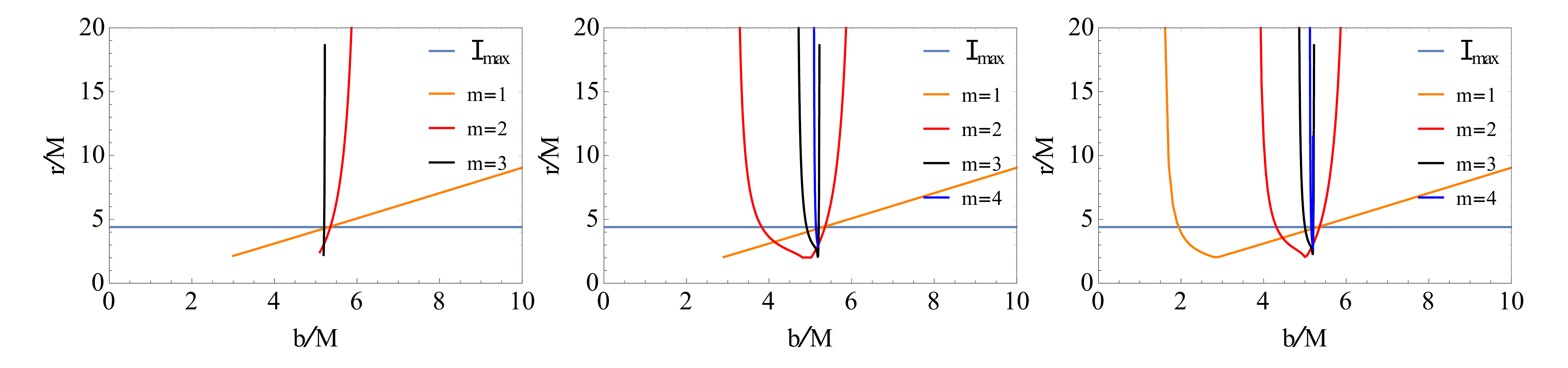}
		\caption{Transfer function $r_m(b)$ for first few  intersections for different models: black hole, transparent gravastar and reflective gravastar separately. The orange line is the first intersection (m=1); the red line is the second intersection (m=2); the black line is the third intersection (m=3) and the dark blue line is the forth intersection (m=4). The light blue line, labeled by $I_{\rm peak}$, corresponds to $r=r_{\rm peak}$, which leads to the maximum intensity that an individual ring can contribute.}
		\label{transfer}
	\end{figure*}

	In order to study the image of a gravastar/black hole under illumination from a disk, we will first need to understand how light rays that enter from infinity gets deflected around the gravastar/black hole.  	Due to spherical symmetry, we only need to consider equatorial null geodesics (or {\it rays}), parametrized by
	\begin{equation}
	x^\mu(\lambda) =( t(\lambda),r(\lambda),\theta = \pi/2,\phi(\lambda))
	\end{equation}
	and we write $u^\mu = \dot x^\mu$, where dot is derivative with respect to $\lambda$.
	Energy and angular momentum conservation leads to
		\begin{equation}
	E =-u_t =f(r) \dot t \,,\quad L = u_\phi = r^2 \dot\phi
	\end{equation} 
	and the null geodesic condition then leads to 
	\begin{equation}
	\frac{d\phi}{dr} =\pm \frac{1}{r^2}\left[\frac{1}{b^2}-\frac{f(r)}{r^2}\right]^{-1/2}
	\label{geodesic}
	\end{equation}
	with the same $f(r)$ given by Eq. \ref{fcondi}. Here $b$ is the impact parameter for incoming null rays at infinity; we shall always be considering $\phi=0$ when $r\rightarrow +\infty$.
		It is convenient to define an effective potential,
	\begin{equation}
	    V_{\rm eff} = \frac{f(r)}{r^2}\,.
	\end{equation}
	Note that in the Schwarzschild case, this potential peaks at $r=3M$, and this is often referred to the {\it light ring}, the location of the unique circular orbit for photons.  For gravastars, the value of $V_{\rm eff}$ is modified for $r<a$.  The modified potential rises up to infinity at $r=0$.   This means a ray governed by this effective potential will always be ``repelled'' by the origin of $r=0$ --- corresponding to the ray entering the gravastar, grazing around the origin, and eventually returning to the exterior of the star, ending up at infinity again.

	In the black hole case, we have a {\it critical impact parameter} of 
	\begin{equation}
	    b_{\rm crit}=3\sqrt{3}M \approx 5.19M\,.
	\end{equation}
	Rays with $b>b_{\rm crit}$ will be deflected and leave the black hole, while those with $b<b_{\rm crit}$ will be captured by the black hole.  In this way, as we trace a ray that enters from infinity, if $b < b_{\rm crit}$, the tracing ends as the ray reaches the horizon at a finite value of $\phi$, while if $b>b_{\rm crit}$, we will first trace the ray to a position where $r$ reaches the minimum $r_{\rm min}$, satisfying
	\begin{equation}
	\label{eq:rmin}
	    \frac{1}{b^2} = V_{\rm eff}(r_{\rm min})\,,
	\end{equation}
	and then trace its trip back to infinity again using Eq.~\ref{geodesic}.
	
	For gravastars, as long as $a<3M$, the location of the peak of the Schwarzshild effective potential is the same. As a consequence, if $b > b_{\rm crit}$, the geodesic will be the same for both transparent and reflective gravastar as it for black hole: it reaches $r_{\rm min}$ given by Eq.~\eqref{eq:rmin}, and then returns to infinity. 
	
	For $b<b_{\rm crit}$, the ray will reach the surface of the gravastar, and at this point, we would like to consider {\it two different scenarios}: the transparent, and the reflective gravastar. 
	
		\begin{table}[t]
		\centering
		\begin{tabular}{c|cc|cc|cc|}
			&  $b^{\rm BH}_{\rm min}$ & $b^{\rm BH}_{\rm max}$ & $b^{\rm trans}_{\rm min}$ & $b^{\rm trans}_{\rm max}$ & $b^{\rm refl}_{\rm min}$ & $b^{\rm refl}_{\rm max}$\\
			\hline 
			$m=1$ & $2.9M$ & $\infty M$ & $2.9M$ & $\infty M$ & $1.6M$ &  $\infty M$\\ 
			$m=2$ & $5.1M$ & $6M$ & $3.22M$ & $6.02M$ & $3.85M$ & $6.02M$\\ 
			$m=3$ & $5.19M$ & $5.22M$ & $4.69M$ & $5.22M$ & $4.84M$ & $5.22M$\\
			$m=4$ & - & - & $5.08M$ & $5.19M$ & $5.11M$ & $5.19M$\\
			$m=5$ & - & - & - & - & $5.17M$ & $5.19M$
		\end{tabular}
		\caption{The range of $b$ for $m$-th intersection of null ray with disk. For gravastars, $a=2M$ in these cases.}
		\label{tab:rbvalue}
	\end{table}

    The {\it transparent} gravastar has been considered by Ref.~\cite{sakai2014gravastar}. Such a gravastar allows the null ray to pass through its surface, continuing into the interior,  de-Sitter region, which has an effective potential corresponding to 
    \begin{equation}
        f(r)=1-\frac{8\pi \rho r^2}{3}\,.       
    \end{equation}
    As discussed by Ref.~\cite{sakai2014gravastar}, the subsequent trajectory in the interior region appears as a straight line in the $(r\cos\theta,r\sin\theta)$ Cartesian coordinate system. It is straightforward to follow this straight line through the gravastar interior, then follow the subsequent trip back to infinity.  The quantity $d\phi/dr$ is continuous across the gravastar surface upon transmission, due to continuity of $f$ and conservation of angular momentum. One such ray is illustrated in the middle panel of Fig.~\ref{compareorbit}.
    
    For {\it reflective} gravastar, the null ray will neither penetrate or be captured; instead, it will be bounced back at the surface of gravastar, with the same $d\phi/dr$. The ray then returns to infinity following a trajectory that is identical to the incoming one, but reflected.  One such ray is illustrated in the right panel of Fig.~\ref{transfer}. 
	
	We notice that when $a\rightarrow 2M$, even though the time $t$ that takes by the ray to reach $r=a$ approaches infinity, the trajectory of the null ray in the $\phi$-$r$ space approaches the same.  In this way, as we study the shadow of extremely compact gravastars, we can simply take $a=2M$ in our calculations. 
	
	\subsection{Geodesics and Disk Emission Model}
	\label{sec:shadow} 
	
    In order to compute the image of the gravastar under illumination from disk emission, we will follow the same strategy as Ref.~\cite{gralla2019black}, and trace parallel null rays that enter from infinity, {\it back toward}  the vicinity of the gravastar.  (We shall denote that impact parameter of the ray as $b$.)  	We shall focus on a transparent disk in this paper, assuming that it is located  at $r>2M$, and $\phi=\pi/2$, as well as $\phi = 3\pi/2$, as has been indicated as orange line in Figure.~\ref{compareorbit}. Whenever the backward traced ray  intersects with the disk, $\Delta \phi = (2m-1) \cdot \pi / 2$ where $m = 1,2,3...$, it will picks up additional intensity. After accounting for all the intersections, the overall intensity viewed by observer at infinity, corresponding to impact parameter $b$,  can be written as:
	\begin{equation} 
	I(b)=\sum_m g^4 I(r) \bigg |_{r=r_m(b)}
	\label{eq:iobs}
	\end{equation}
	Here 
	\begin{equation}
	    g= \sqrt{\frac{1-2M}{r}}
	\end{equation}
	is the redshift factor.

	\begin{figure}[htb]
	    \centering
	    \includegraphics[width=6cm]{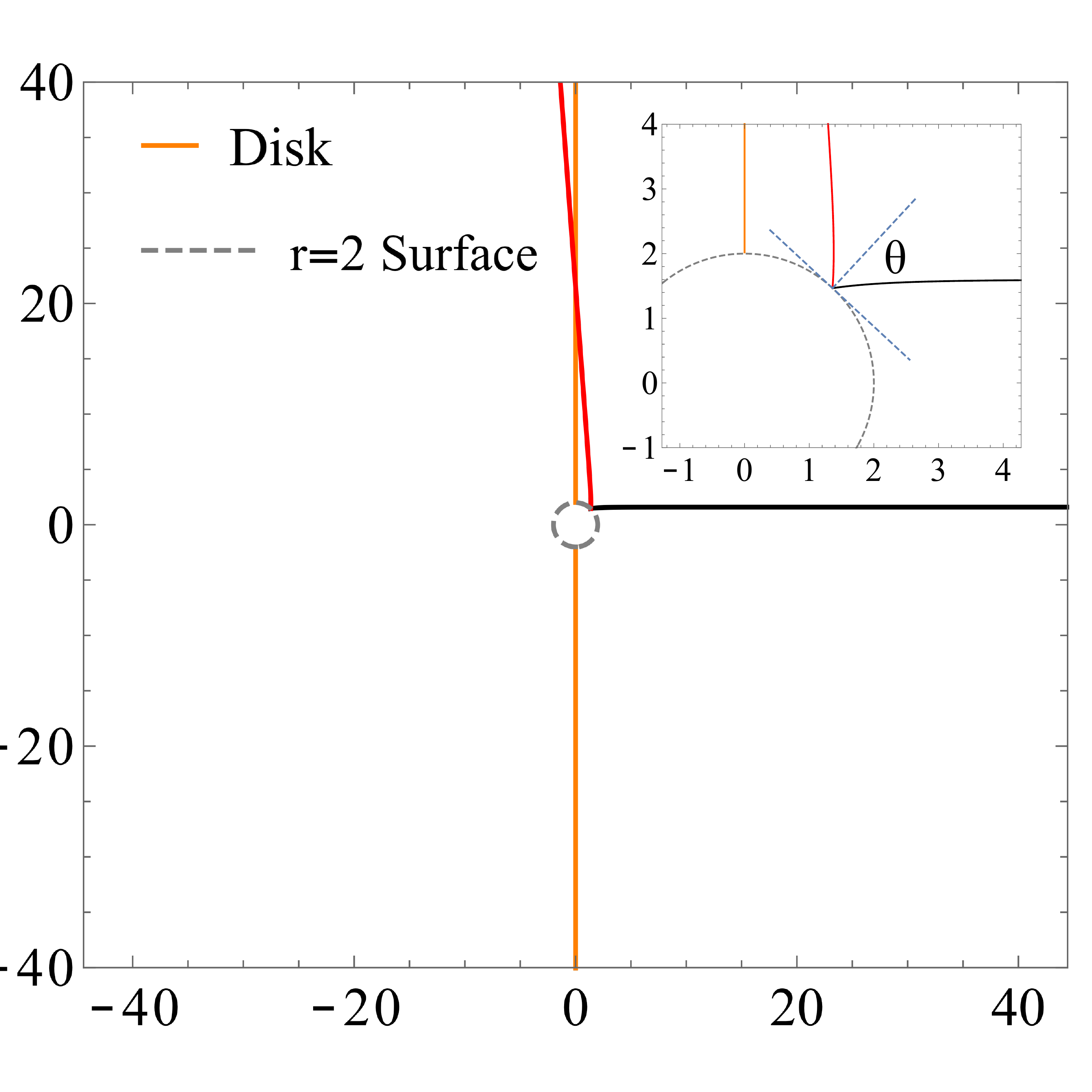}
	    \caption{The null ray orbit for reflective gravastar with $a=1.6M$. The reflected null ray will still intersect with disk at large radius. The small panel shows a detailed graph at the reflected point. The reflection angle $\theta$ shown in the graph is not small enough to fully bounce back the null ray.}
	    \label{orbitrefeg}
	\end{figure}

	Here $r_m(b)$ is a {\it transfer function} that indicates the value of $r$ when the null ray with impact parameter $b$ intersects the disk at the $m$-th time. Fig.~\ref{transfer} shows the first few $r_m(b)$ for the black hole and gravastar spacetimes.  Let us discuss features of these $r_m(b)$ in the following. 
	
	Transfer functions for the first intersection, $r_1(b)$ are shown as orange curves in Fig.~\ref{transfer}.  The black-hole transfer function (left panel) has a domain of $b > 2.9M$; for  $b<2.9M$, the ray goes directly to the horizon, without intersecting the disk. For the transparent gravastar (middle panel),  $r_1(b)$ is the same as it for black hole because the incoming ray is already inside gravastar when it reaches the intended position of the disk and thus have no intersection with the disk. For reflective gravastar (right panel), $r_1(b)$ has an abrupt increase at $b\sim 1.6M$. At such impact parameters, the {\it incoming} ray will hit the gravastar at $r\approx 2M$ and be reflected by an angle less than, but approaching $\pi/2$; therefore, the {\it returning} ray will still intersect with the disk at a large radii. This is illustrated in Fig. \ref{orbitrefeg}.
	
	Transfer functions for the second intersection, $r_2(b)$ are shown as red curves in Fig.~\ref{transfer}.
	For a black hole, $r_2(b)$ exists for $5.1<b/M<6.02$, and is a monotonically increasing function of $b$.  This is understandable, because this intersection takes place when the ray turns almost $\pi$ angle from the initial ray. As $b$ increases, the deflection angle decreases, leading to the intersection taking place at larger $r$.  However, for both transparent and reflective gravastars, the support of $r_2(b)$ is further extended to smaller values of $b$, less than $5.1M$.  At these angles, the rays would hit the black hole and get absorbed, while they can either penetrate or reflect off the gravastar, allowing them to intersect the disk an additional time at $\phi=(3/2)\pi$ as seen in Fig. \ref{compareorbit}.
	
	For the third intersection ($m = 3$), the $r_3(b)$ trajectory will almost be a vertical line in all situations. In addition, the gravastar will even has the fourth and fifth intersection; however, as it will turn out, the intensity produced by these intersection is negligible and we show the approximate domain of these intersection is shown in Table. \ref{tab:rbvalue}
	
\begin{figure}[htb]
	    \centering
	    \includegraphics[width=0.5\textwidth]{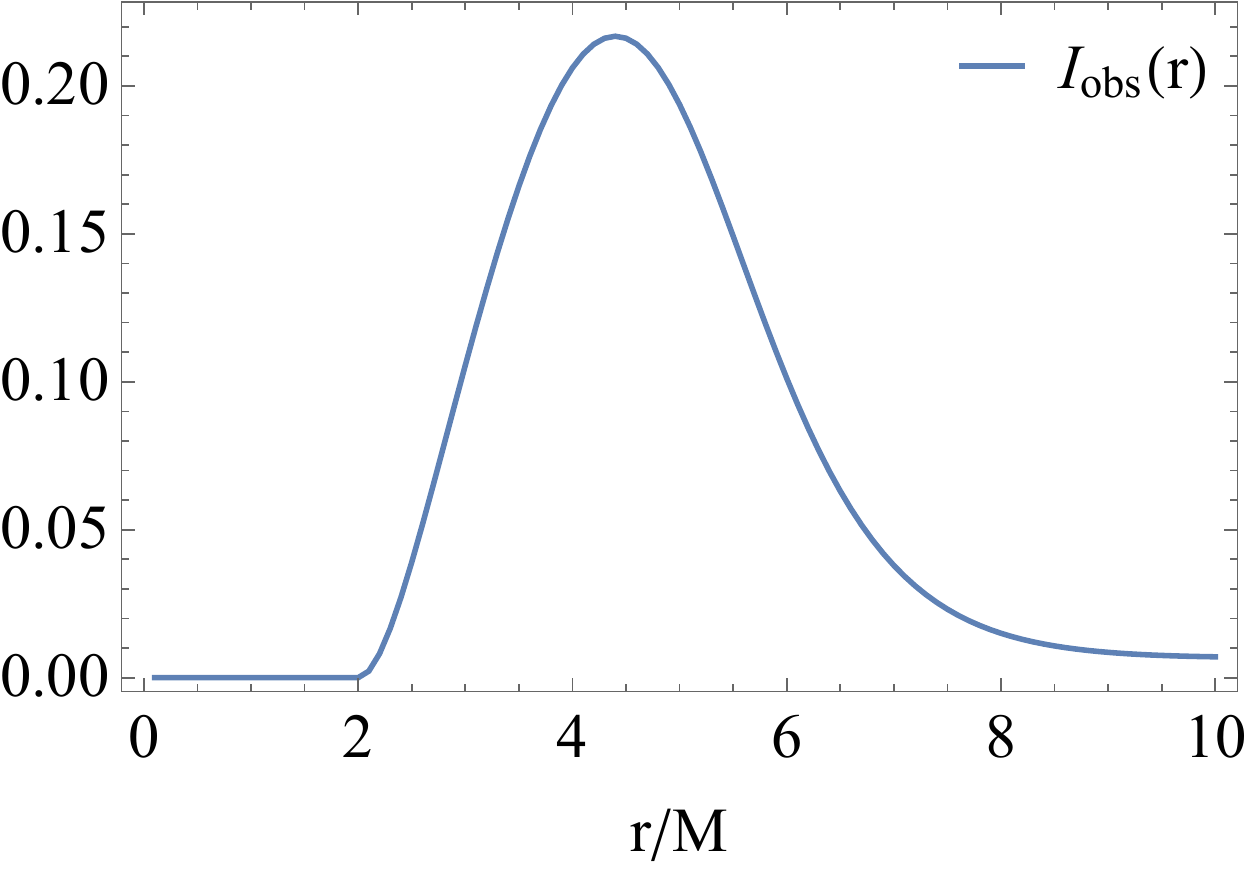}
	    \caption{The plot of Eq. \ref{eq:max}. Both $g^4$ and $I(r)$ depend on the variable $r$. When $r$ is large, the emission is far from the center of the disk, the emission intensity $I(r)$ will be lower while $g^4$ will be larger. The maximum value of $I_{obs}$ exists at $r \sim 4.4M$.}
	    \label{fig:ir}
	\end{figure}

\begin{figure}[t]
	\centering
    	\includegraphics[width=0.8\textwidth]{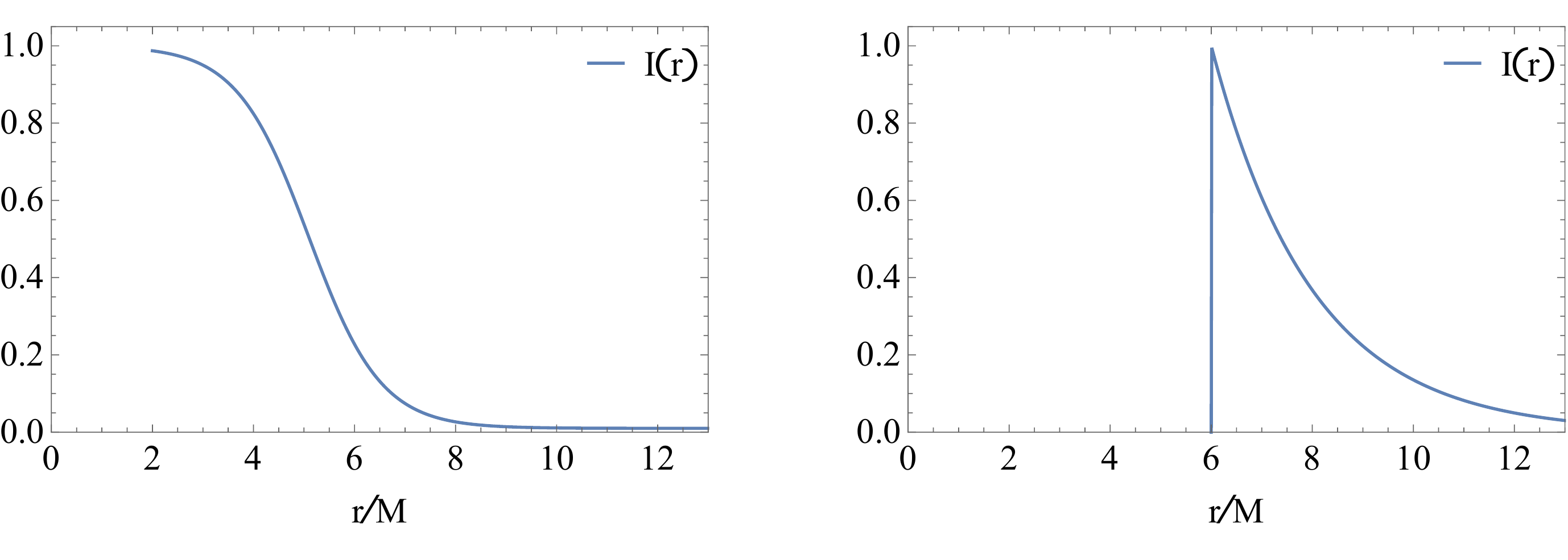}
		\caption{The emission models $I_{\rm em}$ that we will use in the later discussion. Both models are adopted from Ref.~\cite{gralla2019black}.}
		\label{iem}
	\end{figure}

	In this paper, we shall mainly discuss with one of the disk models favored by Ref.~\cite{gralla2019black}, which has emissivity extending all the way toward $r=2M$ --- as is shown in the left  panel of  Fig. \ref{iem}, as well as another disk model, also from Ref.~\cite{gralla2019black}, for comparison. In Fig. \ref{fig:ir}, we also plot 
	\begin{equation}
	    I_{\rm obs}(r)= g^4(r) I_{\rm em}(r)
	    \label{eq:max}
	\end{equation} 
	As we can see from the plot, $I_{\rm obs}$ peaks $r  =r_{\rm peak}\sim 4.4M$.  As we shall see later in the paper, this peak plays a crucial role in the type of rings that we get in the shadow.
	
	\begin{figure*}[t]
	\centering
		\includegraphics[width=0.8\textwidth]{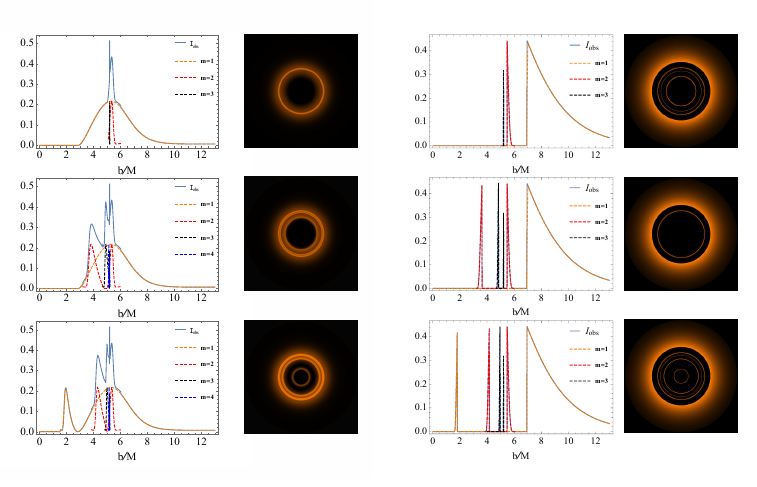}
		\caption{The graphs can be separated into two groups with the left one is plotted with emission model on the left of Fig.~\ref{iem} and the right one is plotted with emission model on the right of Fig.~\ref{iem}. The panels at left of each group show the observed intensity $I_{\rm obs}$ with respect to impact parameter $b$ (light blue line). The colored dash line indicates the intensity produced by the $m$-th intersection where orange is the first intersection ($m=1$), red is the second intersection ($m=2$), black is the third intersection ($m=3$) and dark blue is the fourth intersection ($m=4$). The right panels of each group show the visual emission profile of each circumstance. For two groups, the top panels are plotted for black hole; the middle panels are plotted for transparent gravastar and the bottom panels are plotted for reflective gravastar. The panels at right of each group are the corresponding images of the black hole/gravastar shadows and rings.}
		\label{ibg}
	\end{figure*}
	
	\section{Shadow of the gravastar under different circumstances}
	\label{sec:shadowdiffer}

In this section, we compute the images of black hole and gravastars, and discuss how their features are related to ray propagation. 

\subsection{Computation of Intensity Profile}
\label{sec:3a}

    In Ref. \cite{gralla2019black}, Gralla {\it et al.} classified {\it rings}, e.g., local maximum of the intensity profile, in a black hole's image into two types: {\it photon rings} and {\it lensing rings}. The lensing ring consists of null rays that intersect with the disk two times, while  photon rings consist of null rays that intersect with the disk more than two times. However, this classification is not convenient for our later discussion so that we will define anther two type of rings according to transfer function: {\it incident rings} and {\it deflected rings}. Mathematically, the intensity profile of the image $I(b)$ can be written as a superposition of multiple terms, 
    \begin{equation}
        I(b) = \sum_m I_{\rm obs}(r_m(b))
    \end{equation}
 with the first term, $m=1$, $I_{\rm obs}(r_1(b))$ corresponding to the {\it incident ring}, while subsequent terms $I_{\rm obs}(r_{\rm m}(b))$, $m > 1$, corresponding to the {\it deflected rings}. Each trace in Fig.~\ref{transfer}, illustrates how one particular $r_m$ maps $b$ to the disk, hence picking up intensity $I_{\rm obs}$ (shown in the lower panel of Fig.~\ref{iem}) from the position $r_m(b)$. 

\subsection{Comparisons between Black Hole and Gravastar Images}
\label{sec:compare}

    In the black hole case, we reproduce the result of Ref.~\cite{gralla2019black} in the upper panels of Fig.~\ref{ibg}, but showing intensity contributions from incident and deflected rings separately using different traces in the upper left panel.  The incident ring (orange trance in the left panel) has a smooth and broad distribution that nearly replicates the $I_{\rm obs}$ profile.  The two significant deflected rings are due to $r_2$ (red dashed trace) and $r_3$ (black dashed trace). We get one peak each, for $m=1$ this is because $r_1$ is a monotonic function that only intersects $r_{\rm peak}$ once, while for $m=2$ this is because $r_2$ only has support for $5.19M<b<5.22M$, a very narrow region.

    The middle and bottom rows of Fig.~\ref{ibg} are plotted for transparent gravastar and reflective gravastars, respectively. Under the same emission model, more peaks show up in these images, in comparison to the black hole case. 
    
    For the transparent gravastar (middle panels), the $m=1$ emission is similar to the black-hole case, while the $m=2$ and $3$ emissions each picks up an additional peak, while an $m=4$ emission also shows up. Even though we can count 6 different peaks from these separate contributions, the superposition of these structures leads to 4 separately visible local maxima in the intensity distribution. Next, let us discuss these separate contributions in detail.  
    
    The $m=2$ contribution (red dashed) now has two peaks (showing up as two rings in the image), at $b/M=3.8$ and $4.91$, respectively --- corresponding to intersections of the $m=2$ trace with the $I_{\rm max}$ trace, in the middle panel of Fig.~\ref{ibg}.  This means, light emitted at $r_{\rm peak}$ by the disk have two different ways of reaching infinity after intersecting the disk itself for one additional time (so in total $m=2$).  The $m=3$ contribution (black dashed) also has two peaks (showing up as two additional peaks in the image), at $b/M=4.92$ and $5.2$, respectively.  Similar to the $m=2$ case, they correspond to the two intersections between the $m=3$ trace and the $I_{\rm max}$ trace in the middle panel of Fig.~\ref{transfer}; light from $r_{\rm peak}$ on the disk has two different ways of reaching infinity, after intersecting the disk for two additional times.  Finally, the $m=4$ contribution is narrowly peaked at $b=5.13M$.

    
    In the case of a reflective gravastar (lower panels of Fig.~\ref{ibg}), the $m=2,3,4$ contributions are qualitatively similar, but there is an additional peak at $b \sim 1.94 M$ -- which is visually very distinctive from images of the black hole and the transparent gravastar.   We can count 7 different peaks from the separate contributions, which lead to 5 visible local maxima in the image.

    We can see that the additional peak at $b \sim 1.94 M$  is caused by the first intersection ($m=1$).  More specifically, in the right panel of Fig.~\ref{transfer}, the $m=1$ trace intersects the $I_{\rm max}$ trace twice -- this is due to the reflection of rays on the gravastar surface, as we had discussed in Fig.~\ref{transfer}.  
    
    In comparison with the emissions (right groups of Fig.~\ref{ibg}) under the emission model on the right panel of Fig.~\ref{iem}, we can see that even with different model, the gravastar will always exhibit more rings than black hole do. For reflective gravastar, there is also an additional ring in near $m\sim1.8M$ which is caused by the first intersection ($m=1$). In general, the gravastar under two different emission models exhibit almost the same pattern in the differences between black hole.

    \subsection{Discussions and general features} 
    \label{sec:3c}

	
	As we have seen above, the locations of peaks in the images correspond to
	values of $b$ for which 
	\begin{equation}
	    r_{\rm m}(b_{\rm ring})=r_{\rm peak}
    \end{equation}
The width of each peak will thus be determined by the slope of each $r_{\rm m}(b)$. In fact, if the width of the emission profile is given by $\Delta r= \sigma_{\rm em}$, then the width of the peak, at least when the peak is narrow, can be approximated by 
\begin{equation}
    \Delta b = \frac{\sigma_{\rm em}}{r'_m(b_{\rm ring})}
    \label{eq:slope}
\end{equation}
where $r'_m$ is the derivative of $r_m$. In this way, traces of $r_{\rm m}$ that intersects the $r_{\rm peak}$ with a small slope will lead to a wider peak in the image, while a large slope will lead to a narrower peak.

Besides the two types of gravastar we have discussed above, there is also a situation that the gravastar can absorb the photon, which makes the null ray neither go through the gravastar nor be reflected by it. As we have discussed above, the geodesics for null ray are the same when it is outside either gravastar or black hole; therefore, the black hole and gravastar will exhibit the same features under this situation as both of them absorb the photon.
	\section{Conclusions}
	\label{sec:conclusion}
	
	In this paper, we have discussed null geodesic propagations in both transparent and reflective gravastar spacetimes, and obtained predictions for images of these objects when illuminated by a thin, transparent accretion disk.  We have restricted ourselves to the simple situation, in which the observation is done ``face-on'': if the disk is in the $x$-$y$ plane, then the observer at infinity is located along the $z$ axis.  We have also focused on gravastars that are ultracompact. 
	
	The null geodesic will be the same for both type of gravastars and black hole when impact parameter $b > b_{\rm crit}$, where $b_{\rm crit} = 3\sqrt{3}M \sim 5.19 M$. However, the null geodesic will propagate differently when $b < b_{\rm crit}$ --- in particular for those rays, which, in the black hole case, would have been captured by the horizon. For transparent gravastars, the rays will traverse the de-Sitter interior (along a straight line in the coordinate system adopted by our paper, see Fig.~\ref{compareorbit}), while for reflective gravastars, the rays will bounce on their surfaces. 
	
	We then used the ray-tracing prescription and disk emission model by Gralla {\it et al.}, and showed that these new ways of propagation lead to qualitatively different images.  Thanks to the simplicity of the Gralla {\it et al.} model, contribution of the disk to observed intensity peaks at a location of $r=r_{\rm peak}$. This leads to a simple ring structure in the image. As discussed in Sec.~\ref{sec:compare}, the rings will occur at values $b$ where $r_{\rm m}(b) = r_{\rm peak}$ while the width of the ring is depends on the slope   of each $r_{\rm m}(b)$, see Eq.~\ref{eq:slope}. 
	
	More specifically, using this simple disk model, we showed that the transparent gravastar will have two additional ``rings'' in their image, while the reflective gravastar will have three additional rings --- with one that is half way between the center and the other rings.   
	
	The double rings image also appeared in other studies. In black hole situations, Vincent {\it et al.}\cite{Vincent2020} presented that the additional rings, or ``secondary ring'' from their definition, are  formed not only by the gravitation, but also have relation with the accretion-flow. Besides black hole, in wormhole situation studied by Wielgus {\it et al.}\cite{Wielgus2020}, the additional ring can also form by the null rays that travel to the other side of wormhole and go back, which thus form another ring.
	
	At the end, let us emphasize that, even though we have only considered a face-on viewing angle for a non-spinning gravastar, unlike other studies we discussed above, the way the innermost incident ring shows up for the reflective gravastar is very robust. For this reason, we think images of M87 can be used to probe the reflectivity of a hypothetical gravastar at its center --- or to set an upper bound for the reflectivity of the black hole.

	\section*{Acknowledgements}
	The author would like to thank Yanbei Chen for suggesting this research project, help on technical issues, and detailed comments on the manuscript.

	\bibliographystyle{ws-ijmpd}
\bibliography{references}

\end{document}